\documentclass[preprint,12pt]{elsarticle}
\usepackage{graphicx}
\usepackage{amssymb}
\usepackage{amsthm}
\usepackage{lineno}
\usepackage[pdftex]{color}
\usepackage{array}
\usepackage{amsmath}

\journal{Elsevier}

\begin{document}
	
	\begin{frontmatter}
		
		\title{A Comparative Study of Structural Representations for 
			2D Materials: Insights from Dynamic Collision Fingerprint
			and Matminer}
		
		\author{Raphael M. Tromer}
		\address{Institute of Physics, University of Brasília, 70910-900 Brasília, Federal District, Brazil} 
		\author{Isaac M. Felix}
		\address{Center for Agri-food Science and Technology, Federal University of Campina Grande, 58840-000 Pombal, Paraíba, Brazil}
		\author{Rafael Besse}
		\address{International Center of Physics, Institute of Physics, University of Brasília, 70910-900 Brasília, Federal District, Brazil}
		\author{Marcelo L. Pereira Junior}
		\address{Department of Electrical Engineering, College of Technology, University of Brasília, 70910-900 Brasília, Federal District, Brazil} 
		\author{Marcos G. E. da Luz}
		\address{Department of Physics and Multidisciplinary Laboratory for Modeling and Analysis of Data in Complex Systems (MADComplex), Center for Scientific Modeling and Computing, Federal University of Paraná, 81531-980 Curitiba, Paraná, Brazil}

		\begin{abstract}
			In materials science, the selection of structural 
			descriptors for machine learning protocols strongly 
			influences predictive performance and the degree of physical interpretability that can be achieved from the derived models. 
			Although more complex descriptors may improve numerical 
			accuracy, they often represent extra computational load,
			also reducing transparency into the underlying structural 
			information.
			A framework called the Dynamic Collision Fingerprint
			(DCF) was recently proposed with the goal of producing 
			concise, physically significant representations, generating
			descriptors via dynamical probing of atomic structures.
			In this work, we benchmark DCF using a dataset composed 
			of 120 two-dimensional carbon allotropes and compare 
			its performance with the widely considered Matminer 
			library. 
			The analysis employs three regression models,
			linear regression, decision tree, and XGBoost, 
			evaluated over train and test partitions ranging 
			from 10\% to 90\% and repeated over multiple random 
			seeds in order to characterize statistical variability. 
			The obtained results demonstrate that DCF easily matches 
			Matminer in terms of predicting accuracy across all 
			learning algorithms.
			However, it accomplishes this using descriptors that 
			are significantly lower dimensional, pointing to 
			manageable computing costs.
			Moreover, compared to the rather technical Matminer 
			descriptions, the DCF exhibits considerably clearer 
			physical interpretability.
			These findings suggest that DCF is a significant substitute
			for high-dimensional descriptor libraries as structural representation since it is both computationally flexible 
			and physically grounded.
		\end{abstract}
		
		
	\end{frontmatter}

	\section{Introduction}
	
	Atomic level structural characterization constitutes a central element of materials science and computational chemistry, since it provides the basis for establishing structure property relationships that guide both experimental interpretation and the rational design of novel materials \cite{ghiringhelli2015big,ludwig2019discovery,maffettone2021crystallography,ma2024mlmd,shahzad2024accelerating,song2025new,zeni2025generative,allen2022machine,merchant2023scaling}. Over the last decade, the integration of atomistic simulations, structural descriptors, and machine learning models has significantly transformed high throughput materials discovery \cite{allen2022machine,merchant2023scaling,jain2016computational,kim2018machine,shen2022high,abroshan2023high}. In this context, the definition of descriptors that are simultaneously efficient, interpretable, and transferable has emerged as a critical challenge, since the predictive capability and robustness of machine learning models depend strongly on the quality and representational adequacy of the selected structural features \cite{mobarak2023scope,li2023critical,wang2025evaluating}.
	
	A large variety of structural descriptors has been proposed and consolidated in the literature \cite{bartok2013representing,behler2011atom,rupp2012fast,faber2015crystal,hansen2015machine,chadha2017voronoi,huo2022unified,wang2025graph}. These approaches include representations based on radial and angular distribution functions, orientational order parameters, spectral formulations such as the Smooth Overlap of Atomic Positionsv (SOAP) method \cite{bartok2013representing}, and graph based or topology driven embeddings derived from message passing frameworks \cite{wang2025graph,gilmer2017neural}. Many of these descriptors are available through dedicated software libraries, among which Matminer \cite{ward2018matminer} has become widely adopted in computational materials modeling due to its extensive collection of structural, electronic, and chemical features and its support for reproducible workflows across diverse material classes.
	
	Despite such versatility, the employment of generic 
	high-dimensional descriptor libraries often introduces 
	practical limitations when applied to structurally 
	complex systems, the case of two-dimensional (2D) materials. 
	A first limitation concerns the physical interpretability 
	of many features, as they cannot be directly associated 
	with intuitive structural characteristics.
	A second limitation concerns sensitivity to disorder,
	defects, and aperiodicity, which are frequently present 
	in 2D systems, where local distortions, vacancies, 
	and topological irregularities are common \cite{pereira2008modeling,hong2015exploring,thomas2022point,kirchhoff2022electronic}.
	
	Motivated by these challenges, a recently proposed 
	descriptor scheme, the Dynamic Collision Fingerprint (DCF), 
	was introduced aiming to be a more physically based 
	alternative \cite{tromer2025dynamic}. 
	Guided by well-established concepts from classical statistical mechanics \cite{drude1902elektronentheorie,schneider-1983}, 
	DCF departs from static coordinate-based representations.
	Instead, it seeks to dynamically probe atomic structures 
	through trajectories of idealized particles undergoing 
	elastic collisions with the lattice. 
	In this way, statistical analyses of free paths, collision
	angles, recurrence events, and angular symmetries obtained 
	through Fourier analysis and Shannon entropy allow the
	descriptor to encode structural signatures associated with 
	symmetry, porosity, and disorder \cite{tromer2025dynamic}. 
	This construction yields descriptors that are directly 
	interpretable in physical terms, independent of electronic 
	or energetic models, and computationally adjustable 
	according to the assumed sampling parameters. 
	
	Notwithstanding the promising performance of DCF 
	displayed in its initial work \cite{tromer2025dynamic}, 
	a thorough evaluation against popular high-dimensional
	descriptor libraries and under controlled and statistically 
	consistent benchmarking settings is still missing.
	In this contribution we fulfill this gap by 
	systematically contrasting the DCF with the Matminer 
	descriptor library.
	Using a dataset composed of 120 distinct 2D carbon 
	allotropes, descriptors generated by both approaches are 
	considered within three representative machine learning 
	models, namely, linear regression \cite{hope2020linear}, 
	decision trees \cite{quinlan1986induction}, and XGBoost \cite{chen2016xgboost}.
	Also, a wide variety of train and test partitions are examined
	in the comparisons.
	
	The comprehensive analyses carried out in the present
	work allows to evaluate 
	the DCF's relative performance in terms of physical interpretability, computing efficiency, prediction accuracy, 
	and resilience.
	The obtained results clearly demonstrate that the DCF can
	be a complementary or even a substitute alternative 
	descriptor scheme for materials informatics workflows.
	In particular, by reframing structural characterization 
	as a dynamical response problem rather than a purely static geometric picture, we unveil that predictors
	can be fairly compact (low-dimensional), while retaining 
	the information needed to accurately describe important 
	aspects of 2D materials.

	\section{Methodology}
	
\begin{figure}[t!]
		\centering
		\includegraphics[width=0.45\linewidth]{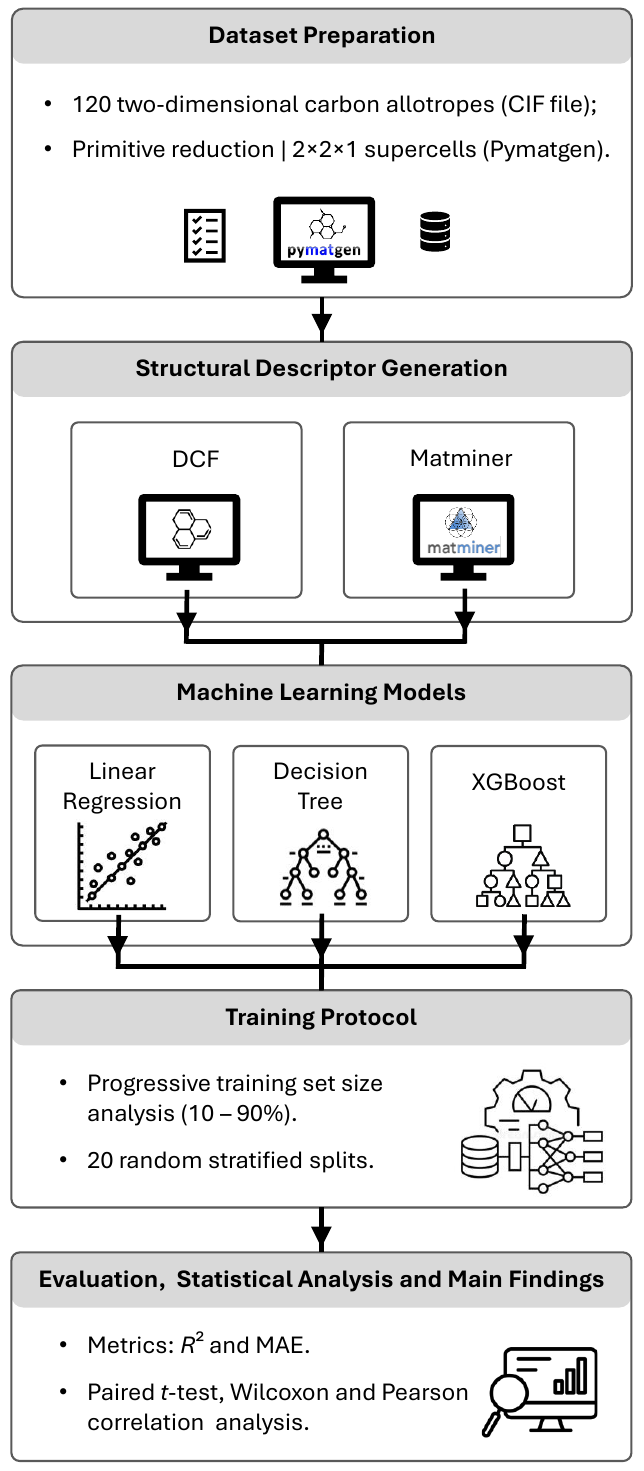}
		\caption{Schematic workflow of the computational pipeline employed in this study. 
			After dataset preparation, structural descriptors were generated using DCF and Matminer, and the models were trained using linear regression, decision trees, and XGBoost. 
			Performance was assessed through MAE and $R^2$, complemented by paired statistical tests and correlation analyses.}
		\label{fig:fig1}
	\end{figure}

	The methodology followed here comprises dataset 
	preparation, descriptor construction, machine learning 
	modeling, and statistical evaluation, all performed under 
	consistently specified conditions to ensure direct 
	comparability between descriptor approaches. 
	A schematic overview of the complete workflow adopted in 
	this study is presented in Figure~\ref{fig:fig1}, summarizing 
	the main stages from dataset preparation to descriptor 
	generation, model training, and statistical validation.
	
	The benchmark dataset employed in this study comprises
	120 distinct 2D carbon allotropes originally reported 
	in literature \cite{shi2021high}. 
	All coordinates and lattice information for the addressed
	structures are provided in the Supporting Information
	(SI). 
	The systems were standardized using the Pymatgen library 
	\cite{ong2013python}. 
	Each structure was reduced to its primitive cell, 
	assuming a symmetry tolerance of $10^{-3}$~\AA, and 
	atomic coordinates were scaled to ensure uniform density 
	normalization. 
	The chosen target property is the formation energy 
	reported in the literature \cite{shi2021high}. 
	Then, supercells of at least $2\times 2\times 1$ are 
	required for appropriate simulations, and this was 
	consistently adopted throughout the study 
	\cite{tromer2025dynamic}.
	
	Two descriptor frameworks were considered, DCF and 
	the Matminer feature library. 
	The DCF model is explained in detail in 
	\cite{tromer2025dynamic}, but its essential points can 
	be summarized as follows.
	One considers the propagation of classical point 
	particles undergoing elastic collisions within an atomic
	lattice supercell under periodic boundary conditions. 
	Each trajectory propagates during a specific number 
	of steps $N_\text{S}$.
	A single ``step'' corresponds to a straight line segment
	between successive boundary crossings.
	For proper averages, a total of $N_\text{L}$ trajectories 
	are launched. 
	Along propagation, quantities such as traveled distances
	and angular deflections are recorded. 
	The final descriptor vector is constructed from statistical
	analyses of angular distributions using Shannon entropy
	as well as Fourier decomposition of recurrence frequencies 
	(associated with one-to-nine-fold lattice symmetries),
	mean free paths, and recurrence times,
	with explicit expressions provided in \cite{tromer2025dynamic}.
	
	Unless otherwise explicitly stated, the standard 
	parameterization was $N_\text{S}=10^4$ and $N_\text{L}=200$, 
	yielding descriptor vectors from 25 to 30 dimensions. 
	Larger $N_\text{S}$ and $N_\text{L}$ were considered
	for sensitivity analysis, with the fast (extended)
	configuration corresponding to $N_\text{S}=10^3$ and
	$N_\text{L}=100$ ($N_\text{S}=10^4$ and $N_\text{L}=300$). 
	The implementation relies on NumPy, SciPy, and pandas and 
	is available in the repository indicated in this work.
	
	Matminer descriptors were generated from the Matminer 
	library \cite{ward2018matminer}. 
	The descriptor set includes radial distribution functions, 
	binned up to 20~\AA\ and with a resolution of 0.1~\AA, 
	packing density and volume per atom statistics, and 
	basic stoichiometric attributes. 
	Depending on the adopted radial distribution function 
	discretization, the resulting feature vectors contain 
	approximately 200-500 components, thus substantially 
	larger than those of DCF. 
	All descriptors were normalized using $z$ score scaling 
	prior to training.
	
	Three representative regression algorithms were 
	addressed. 
	Linear regression was implemented using ordinary least 
	squares from scikit learn. 
	The decision tree model was trained with a maximum depth 
	of 8 and a minimum sample count per leaf of 2. 
	The XGBoost model was implemented with the hyperparameters 
	n\_estimators, max\_depth, learning\_rate, and subsample 
	set to 500, 8, 0.05, and 0.8, respectively. 
	Prior to training, all features were standardized to a 
	mean of 0 and a variance of 1. 
	No feature selection or dimensionality reduction was 
	applied in order to preserve direct comparability between 
	descriptor schemes. 
	Fixed hyperparameters were adopted for all models, as the 
	objective here is to compare descriptor families under 
	identical modeling conditions rather than to optimize 
	individual performance.
	
	Predictive robustness was assessed through repeated 
	random train and test partitions, where the test 
	fraction $X_\text{T}$ denotes the portion of the dataset
	assigned to the test set.
	We considered $X_\text{T}$ values ranging from 0.1 to 0.9 
	in increments of 0.1.
	Each partition was repeated 20 times with random seeds
	to ensure statistical reliability, and reported quantities
	correspond to averages across all seeds.
	
	Predictive performance was quantified using the 
	coefficient of determination $R^2$ and the mean absolute
	error (MAE), where MAE corresponds to the average absolute
	difference between predicted and reference values. 
	To evaluate whether DCF and Matminer exhibit statistically
	distinguishable performance and to quantify their 
	concordance across test fractions, paired $t$ tests, 
	Wilcoxon signed rank tests, and Pearson correlation 
	analyses were performed on the resulting metric 
	distributions. Statistical analyses were carried out 
	using SciPy, adopting a significance threshold of $p<0.05$.
	
	\section{Results and Discussion}
	
\begin{figure}[t!]
		\centering
		\includegraphics[width=0.5\linewidth]{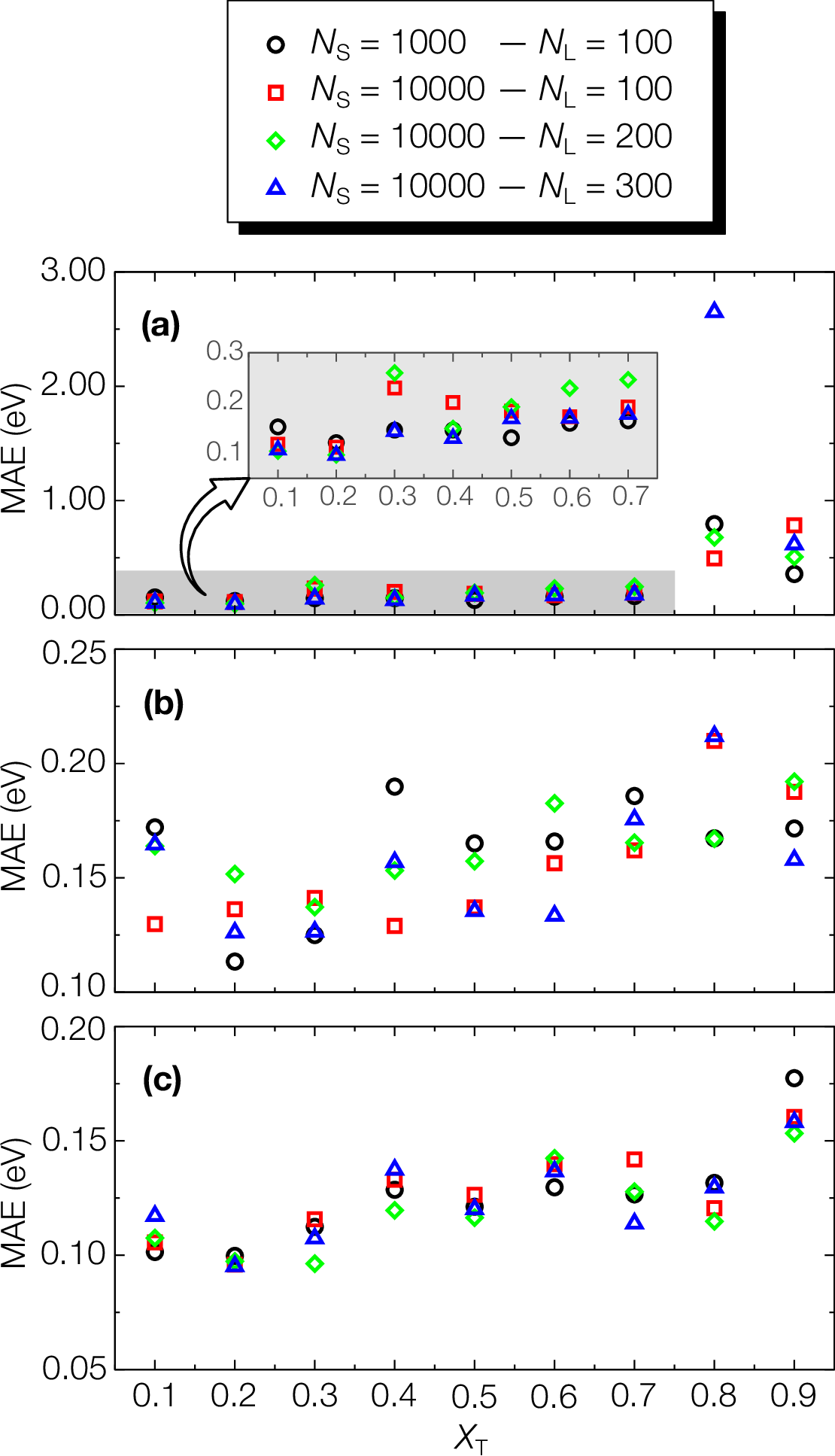}
		\caption{MAE as a function of $X_\text{T}$ for 
			different DCF parameterizations defined by $N_\text{S}$ 
			and $N_\text{L}$ for (a) linear regression, 
			(b) decision tree, and (c) XGBoost.}
		\label{fig:fig2}
	\end{figure}

	The predictive performance of DCF descriptors was first evaluated with respect to the simulation parameters controlling trajectory sampling. Figure~\ref{fig:fig2} summarizes the dependence of the MAE on $X_\text{T}$ for different combinations of $N_\text{S}$ and $N_\text{L}$ and for the three machine learning models considered in this work.
	
	Across all models, the MAE remains stable with respect
	to variations in $N_\text{S}$ and $N_\text{L}$. 
	As observed in Figure~\ref{fig:fig2}(a), the only
	noticeable deviation corresponds to a single outlier 
	near $X_\text{T}\approx 0.8$ in the linear regression
	case. 
	In contrast, Figure~\ref{fig:fig2}(b) and
	Figure~\ref{fig:fig2}(c) show tightly grouped results
	across the full range of $X_\text{T}$, with differences
	comparable to the expected statistical fluctuations 
	from random train test partitions. 
	This behavior indicates that the DCF descriptors are 
	robust with respect to moderate variations in trajectory 
	length and sampling density, and that convergence is
	achieved without the need for extensive parameter tuning.
	
	This stability aligns with the central concept of the DCF formulation \cite{tromer2025dynamic}:
	structural information should be better captured through 
	aggregated dynamical statistics rather than through 
	static high-dimensional signatures. 
	Once the trajectory sampling explores the relevant 
	structural motifs, the resulting descriptors become 
	statistically stable. In the present dataset, increasing $N_\text{L}$ beyond 100 does not produce systematic gains in predictive accuracy, indicating that relatively inexpensive parameterizations are already sufficient to capture the dominant structural information required for machine learning predictions.
	
\begin{figure}[t!]
		\centering
		\includegraphics[width=0.5\linewidth]{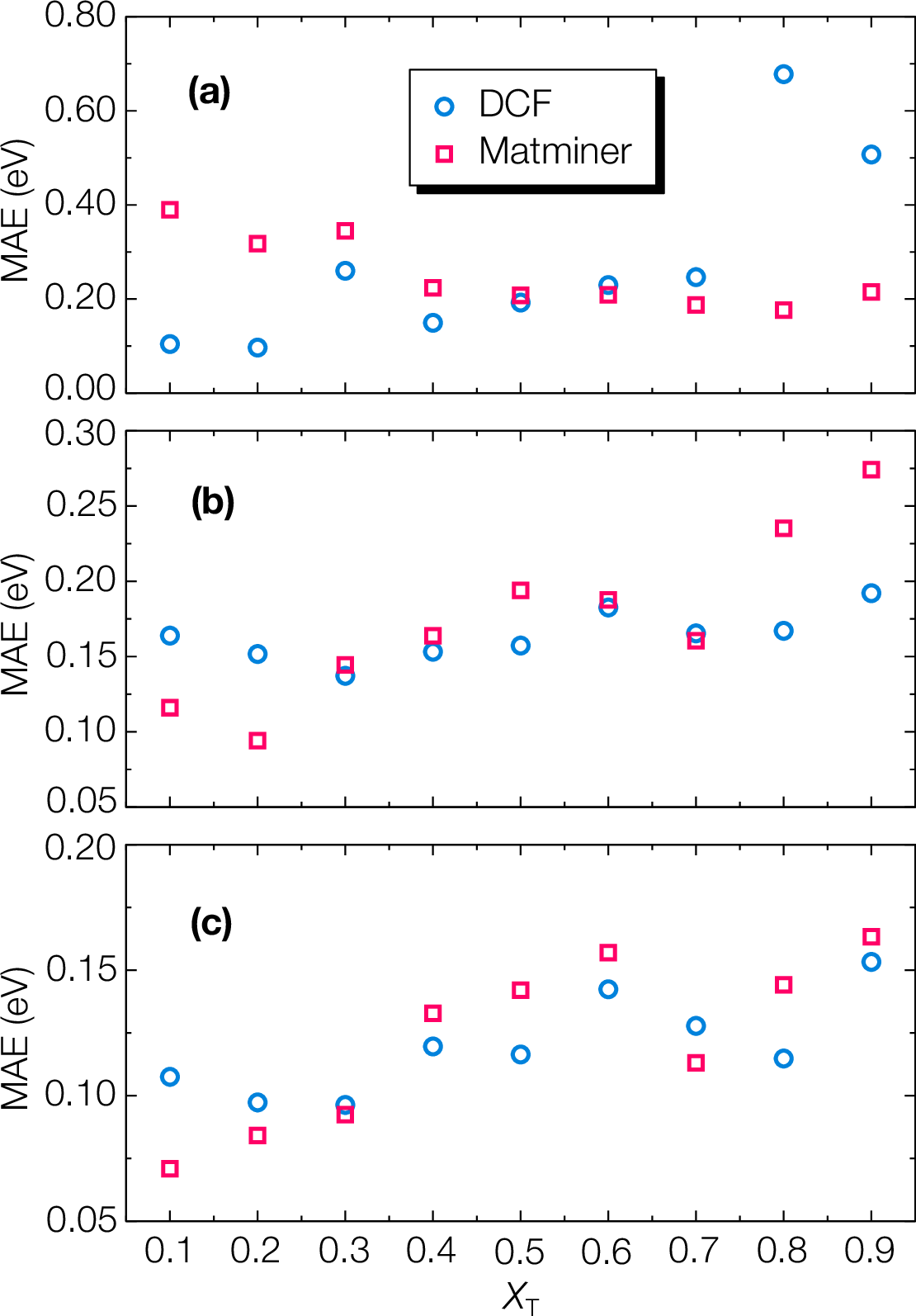}
		\caption{MAE as a function of $X_\text{T}$ comparing 
			DCF and Matminer descriptors for (a) linear regression, 
			(b) decision tree, and (c) XGBoost.}
		\label{fig:fig3}
	\end{figure}

	For the following analyzes, we set $N_\text{S}=10^4$ 
	and $N_\text{L}=200$, representing a balanced trade-off 
	between statistical stability and computational cost. Figure~\ref{fig:fig3} compares the predictive
	performance obtained with DCF and Matminer descriptors 
	through the MAE as a function of $X_\text{T}$ for our
	three machine learning models.
	The results illustrate that DCF systematically reaches 
	predictive errors comparable to those obtained with 
	Matminer, despite its substantially lower dimensionality
	and simpler construction.
	
	For linear regression, depicted in Figure~\ref{fig:fig3}(a),
	DCF frequently produces similar or slightly smaller MAE values across several test fractions, with more pronounced differences emerging at larger $X_\text{T}$, where the reduced amount
	of training data amplifies sensitivity to descriptor quality. In this regime, both approaches occasionally exhibit larger fluctuations, which is expected given the limited expressive capacity of linear models for complex structure property relationships.
	For decision trees, Figure~\ref{fig:fig3}(b) shows that 
	the two descriptor frameworks lead to closely overlapping
	MAE values across most test fractions.
	Also, both representations provide similar levels of 
	structural information for moderate nonlinear learning. 
	
	The comparison becomes even more consistent for XGBoost, 
	as seen from Figure~\ref{fig:fig3}(c), which reveals
	nearly indistinguishable results between DCF and Matminer 
	over the entire range of $X_\text{T}$. 
	This indicates that the essential predictive information
	captured by the higher-dimensional Matminer representation
	is effectively retained within the compact DCF 
	when coupled to a strong nonlinear learner.
	
\begin{figure}[t!]
		\centering
		\includegraphics[width=0.5\linewidth]{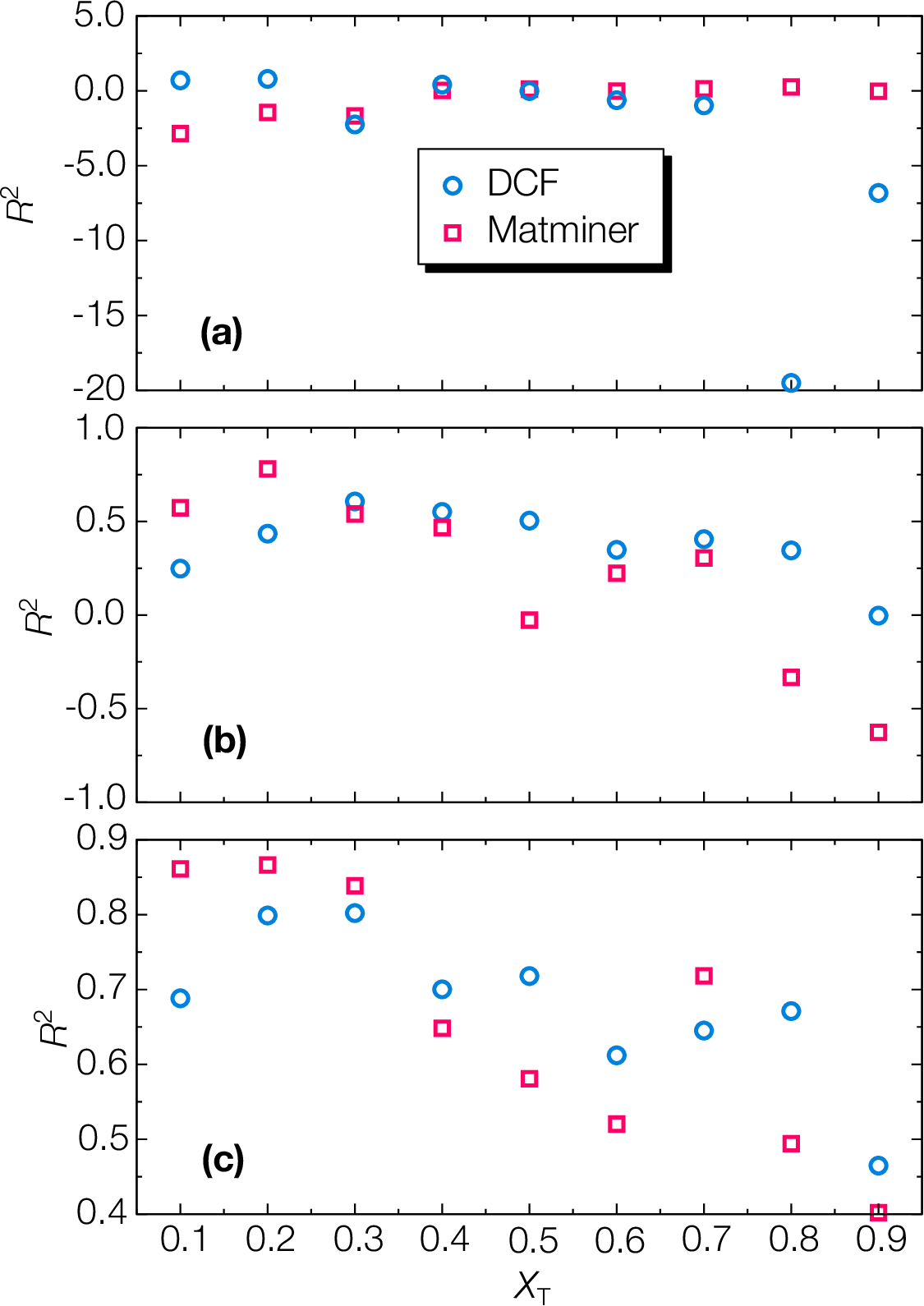}
		\caption{$R^2$ as a function of $X_\text{T}$ 
			comparing DCF and Matminer descriptors for (a) 
			linear regression, (b) decision tree, and (c)
			XGBoost.}
		\label{fig:fig4}
	\end{figure}

	An additional perspective on the predictive capability 
	discussed above is obtained by analyzing the coefficient 
	of determination, whose evolution with $X_\text{T}$ is
	shown in Figure~\ref{fig:fig4} for DCF and Matminer 
	descriptors. 
	We recall that while MAE quantifies the absolute prediction
	error, $R^2$ provides additional insight into how 
	effectively each descriptor framework captures the
	variance of the target property across different train 
	and test partitions.
	
	In the linear regression case, Figure~\ref{fig:fig4}(a), 
	both descriptor sets lead to low predictive power, 
	with $R^2$ values fluctuating around zero and occasionally
	becoming negative at larger $X_\text{T}$. 
	This trait is consistent with the limited flexibility 
	of linear models when applied to complex, intrinsically 
	nonlinear structure-property relationships, reinforcing 
	the interpretation already suggested by the MAE analysis.
	The decision tree model, Figure~\ref{fig:fig4}(b), 
	indicates a clear gain in predictive power, with 
	predominantly positive $R^2$ values across the tested 
	splits. 
	The agreement between DCF and Matminer remains strong, 
	for the small differences observed in specific regions of $X_\text{T}$ not altering the overall picture of comparable performance.

	\begin{table*}[ht!] 
		\centering
		\caption{\textbf{Comparison between DCF and Matminer in terms of feature dimensionality, interpretability, computational cost per structure, and overall predictive behavior.} DCF standard corresponds to $\mathbf{N_\text{S}=10^4}$ and $\mathbf{N_\text{L}=200}$, while DCF fast corresponds to $\mathbf{N_\text{S}=10^3}$ and $\mathbf{N_\text{L}=100}$. Performance summaries are consistent with the trends discussed from Figures~\ref{fig:fig3} and \ref{fig:fig4}. For the Matminer descriptor set, the dimensionality range depends on the selected presets, including RDF binning range and resolution, as well as optional feature subsets.}
		\label{tab:tab1}
		
		\setlength{\tabcolsep}{6pt}
		\renewcommand{\arraystretch}{1.15}
		
		\resizebox{\textwidth}{!}{
			\begin{tabular}{%
					>{\centering\arraybackslash}m{3.2cm}
					>{\centering\arraybackslash}m{2.3cm}
					>{\centering\arraybackslash}m{3.3cm}
					>{\centering\arraybackslash}m{2.3cm}
					>{\centering\arraybackslash}m{6.8cm}}
				\hline
				\textbf{Descriptor set} &
				\textbf{\#Features} &
				\textbf{Interpretability} &
				\textbf{Time per structure} &
				\textbf{Performance summary} \\
				\hline
				DCF (Standard) &
				$\sim$25-30 &
				High (physics based) &
				$\sim$4 min &
				Predictive accuracy comparable to Matminer for both MAE and $R^2$, with consistent performance for decision tree and XGBoost models. \\
				
				DCF (Fast) &
				$\sim$25-30 &
				High (physics based) &
				$\sim$30 s &
				Results remain very close to the standard configuration, with differences within the statistical variations observed in Fig.~\ref{fig:fig2}. \\
				
				Matminer &
				$\sim$200-500 &
				Moderate to low (bin based) &
				$\sim$10 s &
				Comparable MAE and $R^2$ behavior relative to DCF, showing similar trends for decision tree and XGBoost models. \\
				\hline
			\end{tabular}
		}
	\end{table*}
	
	The aforementioned agreement becomes even stronger
	for XGBoost, Figure~\ref{fig:fig4}(c).
	Indeed, the plots show consistently high $R^2$ values 
	for both descriptor frameworks. 
	The close overlap between the results confirms that DCF 
	captures essentially the same predictive information content as Matminer when combined with a sufficiently expressive nonlinear learner. 
	Variations between individual points remain within the range expected from statistical fluctuations across different
	random train-test partitions.
	
	Beyond just the crude face value of predictive metrics, 
	it is also relevant 
	to examine how the two descriptor protocols differ in 
	terms of dimensionality, interpretability, and 
	computational cost, as these aspects directly affect their 
	practical applicability in large-scale materials 
	informatics workflows. 
	A central distinction concerns feature dimensionality and
	physical transparency. The DCF representation remains
	compact, typically containing approximately 25 to 30 
	descriptors, each one directly associated with 
	a physical or geometric characteristic, as the already
	mentioned mean free path, recurrence time, angular entropy, 
	and rotational symmetry intensities from 1 to 9-fold. 
	In contrast, the Matminer representation contains several
	hundred features, largely derived from discretizing the
	radial distribution function across multiple distance 
	bins (up to approximately 20~\AA), complemented by 
	density and packing statistics. 
	While such a high-dimensional description is versatile 
	and broadly applicable, individual components are less 
	intuitive from a physical perspective; note that 
	isolated RDF bins rarely carry a clear standalone 
	interpretation.
	
	A second relevant aspect concerns computational cost. 
	In the current workflow, Matminer requires approximately 
	10~s per structure, whereas DCF in its standard configuration ($N_\text{S}=10^4$, $N_\text{L}=200$) requires 
	approximately 4~min per structure. 
	However, the stability analysis previously discussed shows 
	that the DCF's performance is only weakly sensitive to 
	the trajectory-sampling parameters. 
	Even the fast configuration 
	($N_\text{S}=10^3$, $N_\text{L}=100$) yields MAE values
	very close to those obtained with more extensive sampling, 
	while reducing the average wall time to roughly 30~s 
	per structure. 
	In this regime, the necessary computational work is
	fairly of the same order than that of Matminer, while 
	maintaining comparable predictive accuracy across most
	training and test partitions.
	
	Taken together with the results shown in Figures~\ref{fig:fig3} and \ref{fig:fig4}, these observations highlight a clear trade-off between descriptor compactness, interpretability, and computational efficiency. Linear models remain limited across both descriptor frameworks, whereas decision tree and XGBoost models consistently achieve high predictive performance, demonstrating that DCF preserves the essential structural information despite its reduced dimensionality. A consolidated comparison summarizing dimensionality, interpretability, computational cost, 
	and predictive behavior is presented in Table~\ref{tab:tab1}.
	
	To further complement the performance trends discussed 
	above, additional statistical analyses were conducted to 
	evaluate whether DCF and Matminer exhibit significant 
	differences across the full ensemble of repeated random 
	splits. 
	Paired statistical tests were applied to the metric 
	distributions obtained from 20 random seeds for each
	$X_\text{T}$ value, along with correlation analyses 
	across test fractions.
	This provides a quantitative assessment of
	agreement between descriptor frameworks beyond direct 
	visual comparison.
	
	For linear regression, no statistically significant 
	differences were detected between DCF and Matminer for
	either $R^2$ or MAE.
	Both paired $t$-tests and Wilcoxon signed-rank tests 
	yielded non-significant outcomes ($p > 0.05$), indicating
	that the small variations observed between descriptors
	are compatible with statistical fluctuations rather than
	systematic performance differences. 
	Correlation analyses across test fractions revealed weaker 
	alignment than in nonlinear models, consistent with the 
	higher variability observed in this regime, particularly
	at large $X_\text{T}$'s, when both descriptor sets
	can produce unstable, even negative, $R^2$ values.
	For the decision tree model, paired statistical tests 
	again indicated no relevant distinctions between DCF 
	and Matminer for either MAE or $R^2$ ($p > 0.05$).
	Also, performance trends across test fractions remain
	positively correlated, pointing to similar responses of
	both descriptor frameworks to variations in training-set 
	size. Regimes where performance improves or degrades,
	therefore, tend to coincide for DCF and Matminer,
	corroborating the traits observed in
	Figures~\ref{fig:fig3}(b) and \ref{fig:fig4}(b).
	For XGBoost, paired $t$-tests and Wilcoxon tests show
	that DCF and Matminer are statistically indistinguishable
	across all evaluated metrics ($p > 0.05$), while 
	correlation analyses across test fractions reveal 
	strongly positive alignment for both MAE and $R^2$.
	
	\section{Conclusions}
	
	This work presented a systematic comparison of the 
	DCF and Matminer descriptor frameworks, evaluating 
	their predictive performance across multiple machine
	learning models and train–test data splits.
	The analysis combined direct performance curves, 
	statistical tests, and an explicit comparison of 
	descriptor dimensionality, interpretability, and
	computational cost.
	
	Across the full set of simulations, both descriptors 
	exhibited similar predictive limits under linear regression,
	in which model simplicity dominates the learning behavior 
	and leads to unstable performance at large test fractions.
	In contrast, nonlinear learners consistently displayed
	higher accuracy, as it should be expected given the 
	dataset inherently nonlinear structure–property 
	relationships.
	Actually, for decision tree and XGBoost models, DCF and 
	Matminer converged to closely aligned performance profiles. Statistical tests confirmed the absence of significant
	differences between the two descriptor sets, while correlation analyses across test fractions revealed consistent predictive 
	trends. 
	
	More generally, the present contribution comprehensively 
	illustrated a key trade-off between descriptor compactness, 
	interpretability, and computational efficiency. 
	On one hand, Matminer offers a broad, computationally 
	efficient representation but relies on a high-dimensional 
	feature space that is often less directly interpretable. 
	On the other hand, DCF provides a physically grounded and
	compact descriptor whose components remain closely 
	associated with structural and dynamical characteristics,
	while still achieving competitive predictive performance. 
	Furthermore, the stability of DCF under reduced sampling
	parameters enables substantial reductions in computational
	cost without significant loss of accuracy, reinforcing 
	its practical usefulness.
	These positive features support the use of DCF as an 
	effective and scalable framework for data-driven 
	investigations of structure–property relationships 
	in complex materials systems.

	\section*{Conflicts of interest}
	There are no conflicts to declare.
	
	\section*{Acknowledgements}
	\noindent This work was supported by the National Council for Scientific and Technological Development (CNPq), the Coordination for the Improvement of Higher Education Personnel (CAPES), and the Federal District Research Support Foundation (FAPDF). RMT acknowledges CNPq for financial support (grant no. 307371/2025-5). MLPJ acknowledges financial support from FAPDF (grant no. 00193-00001807/2023-16), CNPq (grants no. 444921/2024-9 and 308222/2025-3), and CAPES (grant no. 88887.005164/2024-00). MGEL acknowledges financial support from CNPq (grants no. 307512/2023-1 and 04577/2021-0) and CAPES (grant no. 88881.311780/2018-00).
	
	\bibliographystyle{elsarticle-num.bst}
	\bibliography{bibliography}

\begin{thebibliography}{10}
\expandafter\ifx\csname url\endcsname\relax
  \def\url#1{\texttt{#1}}\fi
\expandafter\ifx\csname urlprefix\endcsname\relax\def\urlprefix{URL }\fi
\expandafter\ifx\csname href\endcsname\relax
  \def\href#1#2{#2} \def\path#1{#1}\fi

\bibitem{ghiringhelli2015big}
L.~M. Ghiringhelli, J.~Vybiral, S.~V. Levchenko, C.~Draxl, M.~Scheffler,
  \href{https://doi.org/10.1103/PhysRevLett.114.105503}{Big data of materials
  science: critical role of the descriptor}, Phys. Rev. Lett. 114~(10) (2015)
  105503.
\newblock \href
  {https://doi.org/https://doi.org/10.1103/PhysRevLett.114.105503}
  {\path{doi:https://doi.org/10.1103/PhysRevLett.114.105503}}.
\newline\urlprefix\url{https://doi.org/10.1103/PhysRevLett.114.105503}

\bibitem{ludwig2019discovery}
A.~Ludwig, \href{https://doi.org/10.1038/s41524-019-0205-0}{Discovery of new
  materials using combinatorial synthesis and high-throughput characterization
  of thin-film materials libraries combined with computational methods}, npj
  Comput. Mater. 5~(1) (2019) 70.
\newblock \href {https://doi.org/https://doi.org/10.1038/s41524-019-0205-0}
  {\path{doi:https://doi.org/10.1038/s41524-019-0205-0}}.
\newline\urlprefix\url{https://doi.org/10.1038/s41524-019-0205-0}

\bibitem{maffettone2021crystallography}
P.~M. Maffettone, L.~Banko, P.~Cui, Y.~Lysogorskiy, M.~A. Little, D.~Olds,
  A.~Ludwig, A.~I. Cooper,
  \href{https://doi.org/10.1038/s43588-021-00059-2}{Crystallography companion
  agent for high-throughput materials discovery}, Nat. Comput. Sci. 1~(4)
  (2021) 290--297.
\newblock \href {https://doi.org/https://doi.org/10.1038/s43588-021-00059-2}
  {\path{doi:https://doi.org/10.1038/s43588-021-00059-2}}.
\newline\urlprefix\url{https://doi.org/10.1038/s43588-021-00059-2}

\bibitem{ma2024mlmd}
J.~Ma, B.~Cao, S.~Dong, Y.~Tian, M.~Wang, J.~Xiong, S.~Sun,
  \href{https://doi.org/10.1038/s41524-024-01243-4}{{MLMD}: a programming-free
  {AI} platform to predict and design materials}, npj Comput. Mater. 10~(1)
  (2024) 59.
\newblock \href {https://doi.org/https://doi.org/10.1038/s41524-024-01243-4}
  {\path{doi:https://doi.org/10.1038/s41524-024-01243-4}}.
\newline\urlprefix\url{https://doi.org/10.1038/s41524-024-01243-4}

\bibitem{shahzad2024accelerating}
K.~Shahzad, A.~I. Mardare, A.~W. Hassel,
  \href{https://doi.org/10.1080/27660400.2023.2292486}{Accelerating materials
  discovery: combinatorial synthesis, high-throughput characterization, and
  computational advances}, Sci. technol. Adv. Mater.: Meth. 4~(1) (2024)
  2292486.
\newblock \href {https://doi.org/https://doi.org/10.1080/27660400.2023.2292486}
  {\path{doi:https://doi.org/10.1080/27660400.2023.2292486}}.
\newline\urlprefix\url{https://doi.org/10.1080/27660400.2023.2292486}

\bibitem{song2025new}
K.~Song, A.~N.~M. Tanvir, M.~O. Bappy, Y.~Zhang,
  \href{https://doi.org/10.1002/smsc.202300359}{New directions for
  thermoelectrics: A roadmap from high-throughput materials discovery to
  advanced device manufacturing}, Small Sci. 5~(3) (2025) 2300359.
\newblock \href {https://doi.org/https://doi.org/10.1002/smsc.202300359}
  {\path{doi:https://doi.org/10.1002/smsc.202300359}}.
\newline\urlprefix\url{https://doi.org/10.1002/smsc.202300359}

\bibitem{zeni2025generative}
C.~Zeni, R.~Pinsler, D.~Z{\"u}gner, A.~Fowler, M.~Horton, X.~Fu, Z.~Wang,
  A.~Shysheya, J.~Crabb{\'e}, S.~Ueda, et~al.,
  \href{https://doi.org/10.1038/s41586-025-08628-5}{A generative model for
  inorganic materials design}, Nature 639~(8055) (2025) 624--632.
\newblock \href {https://doi.org/https://doi.org/10.1038/s41586-025-08628-5}
  {\path{doi:https://doi.org/10.1038/s41586-025-08628-5}}.
\newline\urlprefix\url{https://doi.org/10.1038/s41586-025-08628-5}

\bibitem{allen2022machine}
A.~E. Allen, A.~Tkatchenko,
  \href{https://doi.org/10.1126/sciadv.abm7185}{Machine learning of material
  properties: Predictive and interpretable multilinear models}, Sci. Adv.
  8~(18) (2022) eabm7185.
\newblock \href {https://doi.org/https://doi.org/10.1126/sciadv.abm7185}
  {\path{doi:https://doi.org/10.1126/sciadv.abm7185}}.
\newline\urlprefix\url{https://doi.org/10.1126/sciadv.abm7185}

\bibitem{merchant2023scaling}
A.~Merchant, S.~Batzner, S.~S. Schoenholz, M.~Aykol, G.~Cheon, E.~D. Cubuk,
  \href{https://doi.org/10.1038/s41586-023-06735-9}{Scaling deep learning for
  materials discovery}, Nature 624~(7990) (2023) 80--85.
\newblock \href {https://doi.org/https://doi.org/10.1038/s41586-023-06735-9}
  {\path{doi:https://doi.org/10.1038/s41586-023-06735-9}}.
\newline\urlprefix\url{https://doi.org/10.1038/s41586-023-06735-9}

\bibitem{jain2016computational}
A.~Jain, Y.~Shin, K.~A. Persson,
  \href{https://doi.org/10.1038/natrevmats.2015.4}{Computational predictions of
  energy materials using density functional theory}, Nat. Rev. Mater. 1~(1)
  (2016) 1--13.
\newblock \href {https://doi.org/https://doi.org/10.1038/natrevmats.2015.4}
  {\path{doi:https://doi.org/10.1038/natrevmats.2015.4}}.
\newline\urlprefix\url{https://doi.org/10.1038/natrevmats.2015.4}

\bibitem{kim2018machine}
K.~Kim, L.~Ward, J.~He, A.~Krishna, A.~Agrawal, C.~Wolverton,
  \href{https://doi.org/10.1103/PhysRevMaterials.2.123801}{Machine-learning-accelerated
  high-throughput materials screening: Discovery of novel quaternary heusler
  compounds}, Phys. Rev. Mater. 2~(12) (2018) 123801.
\newblock \href
  {https://doi.org/https://doi.org/10.1103/PhysRevMaterials.2.123801}
  {\path{doi:https://doi.org/10.1103/PhysRevMaterials.2.123801}}.
\newline\urlprefix\url{https://doi.org/10.1103/PhysRevMaterials.2.123801}

\bibitem{shen2022high}
L.~Shen, J.~Zhou, T.~Yang, M.~Yang, Y.~P. Feng,
  \href{https://doi.org/10.1021/accountsmr.1c00246}{High-throughput
  computational discovery and intelligent design of two-dimensional functional
  materials for various applications}, Acc. Mater. Research 3~(6) (2022)
  572--583.
\newblock \href {https://doi.org/https://doi.org/10.1021/accountsmr.1c00246}
  {\path{doi:https://doi.org/10.1021/accountsmr.1c00246}}.
\newline\urlprefix\url{https://doi.org/10.1021/accountsmr.1c00246}

\bibitem{abroshan2023high}
H.~Abroshan, H.~S. Kwak, A.~Chandrasekaran, A.~K. Chew, A.~Fonari, M.~D. Halls,
  \href{https://doi.org/10.1021/acs.chemmater.3c00561}{High-throughput
  screening of hole transport materials for quantum dot light-emitting diodes},
  Chem. Mater. 35~(13) (2023) 5059--5070.
\newblock \href {https://doi.org/https://doi.org/10.1021/acs.chemmater.3c00561}
  {\path{doi:https://doi.org/10.1021/acs.chemmater.3c00561}}.
\newline\urlprefix\url{https://doi.org/10.1021/acs.chemmater.3c00561}

\bibitem{mobarak2023scope}
M.~H. Mobarak, M.~A. Mimona, M.~A. Islam, N.~Hossain, F.~T. Zohura, I.~Imtiaz,
  M.~I.~H. Rimon, \href{https://doi.org/10.1016/j.apsadv.2023.100523}{Scope of
  machine learning in materials research—{A} review}, Appl. Surf. Sci. Adv.
  18 (2023) 100523.
\newblock \href {https://doi.org/https://doi.org/10.1016/j.apsadv.2023.100523}
  {\path{doi:https://doi.org/10.1016/j.apsadv.2023.100523}}.
\newline\urlprefix\url{https://doi.org/10.1016/j.apsadv.2023.100523}

\bibitem{li2023critical}
K.~Li, B.~DeCost, K.~Choudhary, M.~Greenwood, J.~Hattrick-Simpers,
  \href{https://doi.org/10.1038/s41524-023-01012-9}{A critical examination of
  robustness and generalizability of machine learning prediction of materials
  properties}, npj Comput. Mater. 9~(1) (2023) 55.
\newblock \href {https://doi.org/https://doi.org/10.1038/s41524-023-01012-9}
  {\path{doi:https://doi.org/10.1038/s41524-023-01012-9}}.
\newline\urlprefix\url{https://doi.org/10.1038/s41524-023-01012-9}

\bibitem{wang2025evaluating}
H.~Wang, K.~Li, S.~Ramsay, Y.~Fehlis, E.~Kim, J.~Hattrick-Simpers,
  \href{https://doi.org/10.1039/D5DD00090D}{Evaluating the performance and
  robustness of {LLMs} in materials science {Q}\&{A} and property predictions},
  Digital Discovery 4 (2025) 1612--1624.
\newblock \href {https://doi.org/https://doi.org/10.1039/D5DD00090D}
  {\path{doi:https://doi.org/10.1039/D5DD00090D}}.
\newline\urlprefix\url{https://doi.org/10.1039/D5DD00090D}

\bibitem{bartok2013representing}
A.~P. Bart{\'o}k, R.~Kondor, G.~Cs{\'a}nyi,
  \href{https://doi.org/10.1103/PhysRevB.87.184115}{On representing chemical
  environments}, Phys. Rev. B 87~(18) (2013) 184115.
\newblock \href {https://doi.org/https://doi.org/10.1103/PhysRevB.87.184115}
  {\path{doi:https://doi.org/10.1103/PhysRevB.87.184115}}.
\newline\urlprefix\url{https://doi.org/10.1103/PhysRevB.87.184115}

\bibitem{behler2011atom}
J.~Behler, \href{https://doi.org/10.1063/1.3553717}{Atom-centered symmetry
  functions for constructing high-dimensional neural network potentials}, J.
  Chem. Phys. 134~(7) (2011).
\newblock \href {https://doi.org/https://doi.org/10.1063/1.3553717}
  {\path{doi:https://doi.org/10.1063/1.3553717}}.
\newline\urlprefix\url{https://doi.org/10.1063/1.3553717}

\bibitem{rupp2012fast}
M.~Rupp, A.~Tkatchenko, K.-R. M{\"u}ller, O.~A. Von~Lilienfeld,
  \href{https://doi.org/10.1103/PhysRevLett.108.058301}{Fast and accurate
  modeling of molecular atomization energies with machine learning}, Phys. Rev.
  Lett. 108~(5) (2012) 058301.
\newblock \href
  {https://doi.org/https://doi.org/10.1103/PhysRevLett.108.058301}
  {\path{doi:https://doi.org/10.1103/PhysRevLett.108.058301}}.
\newline\urlprefix\url{https://doi.org/10.1103/PhysRevLett.108.058301}

\bibitem{faber2015crystal}
F.~Faber, A.~Lindmaa, O.~A. Von~Lilienfeld, R.~Armiento,
  \href{https://doi.org/10.1002/qua.24917}{Crystal structure representations
  for machine learning models of formation energies}, Int. J. Quantum Chem.
  115~(16) (2015) 1094--1101.
\newblock \href {https://doi.org/https://doi.org/10.1002/qua.24917}
  {\path{doi:https://doi.org/10.1002/qua.24917}}.
\newline\urlprefix\url{https://doi.org/10.1002/qua.24917}

\bibitem{hansen2015machine}
K.~Hansen, F.~Biegler, R.~Ramakrishnan, W.~Pronobis, O.~A. Von~Lilienfeld,
  K.-R. Muller, A.~Tkatchenko,
  \href{https://doi.org/10.1021/acs.jpclett.5b00831}{Machine learning
  predictions of molecular properties: {Accurate} many-body potentials and
  nonlocality in chemical space}, J. Phys. Chem. Lett. 6~(12) (2015)
  2326--2331.
\newblock \href {https://doi.org/https://doi.org/10.1021/acs.jpclett.5b00831}
  {\path{doi:https://doi.org/10.1021/acs.jpclett.5b00831}}.
\newline\urlprefix\url{https://doi.org/10.1021/acs.jpclett.5b00831}

\bibitem{chadha2017voronoi}
A.~Chadha, Y.~Andreopoulos,
  \href{https://doi.org/10.1109/TMM.2017.2673415}{Voronoi-based compact image
  descriptors: Efficient region-of-interest retrieval with {VLAD} and
  deep-learning-based descriptors}, IEEE Trans. Multimed. 19~(7) (2017)
  1596--1608.
\newblock \href {https://doi.org/https://doi.org/10.1109/TMM.2017.2673415}
  {\path{doi:https://doi.org/10.1109/TMM.2017.2673415}}.
\newline\urlprefix\url{https://doi.org/10.1109/TMM.2017.2673415}

\bibitem{huo2022unified}
H.~Huo, M.~Rupp, \href{https://doi.org/10.1088/2632-2153/aca005}{Unified
  representation of molecules and crystals for machine learning}, Mach. Learn.:
  Sci. Technol. 3~(4) (2022) 045017.
\newblock \href {https://doi.org/https://doi.org/10.1088/2632-2153/aca005}
  {\path{doi:https://doi.org/10.1088/2632-2153/aca005}}.
\newline\urlprefix\url{https://doi.org/10.1088/2632-2153/aca005}

\bibitem{wang2025graph}
A.~Wang, G.~C. Sosso,
  \href{https://doi.org/10.1103/PhysRevE.111.064302}{Graph-based descriptors
  for condensed matter}, Phys. Rev. E 111~(6) (2025) 064302.
\newblock \href {https://doi.org/https://doi.org/10.1103/PhysRevE.111.064302}
  {\path{doi:https://doi.org/10.1103/PhysRevE.111.064302}}.
\newline\urlprefix\url{https://doi.org/10.1103/PhysRevE.111.064302}

\bibitem{gilmer2017neural}
J.~Gilmer, S.~S. Schoenholz, P.~F. Riley, O.~Vinyals, G.~E. Dahl,
  \href{https://proceedings.mlr.press/v70/gilmer17a.html}{Neural message
  passing for quantum chemistry}, in: D.~Precup, Y.~W. Teh (Eds.), Proceedings
  of the 34th International Conference on Machine Learning, Vol.~70 of
  Proceedings of Machine Learning Research, PMLR, 2017, pp. 1263--1272.
\newline\urlprefix\url{https://proceedings.mlr.press/v70/gilmer17a.html}

\bibitem{ward2018matminer}
L.~Ward, A.~Dunn, A.~Faghaninia, N.~E. Zimmermann, S.~Bajaj, Q.~Wang,
  J.~Montoya, J.~Chen, K.~Bystrom, M.~Dylla, et~al.,
  \href{https://doi.org/10.1016/j.commatsci.2018.05.018}{Matminer: {An} open
  source toolkit for materials data mining}, Comput. Mater. Sci. 152 (2018)
  60--69.
\newblock \href
  {https://doi.org/https://doi.org/10.1016/j.commatsci.2018.05.018}
  {\path{doi:https://doi.org/10.1016/j.commatsci.2018.05.018}}.
\newline\urlprefix\url{https://doi.org/10.1016/j.commatsci.2018.05.018}

\bibitem{pereira2008modeling}
V.~M. Pereira, J.~Lopes~dos Santos, A.~Castro~Neto,
  \href{https://doi.org/10.1103/PhysRevB.77.115109}{Modeling disorder in
  graphene}, Phys. Rev. B 77~(11) (2008) 115109.
\newblock \href {https://doi.org/https://doi.org/10.1103/PhysRevB.77.115109}
  {\path{doi:https://doi.org/10.1103/PhysRevB.77.115109}}.
\newline\urlprefix\url{https://doi.org/10.1103/PhysRevB.77.115109}

\bibitem{hong2015exploring}
J.~Hong, Z.~Hu, M.~Probert, K.~Li, D.~Lv, X.~Yang, L.~Gu, N.~Mao, Q.~Feng,
  L.~Xie, et~al., \href{https://doi.org/10.1038/ncomms7293}{Exploring atomic
  defects in molybdenum disulphide monolayers}, Nat. Commun. 6~(1) (2015) 6293.
\newblock \href {https://doi.org/https://doi.org/10.1038/ncomms7293}
  {\path{doi:https://doi.org/10.1038/ncomms7293}}.
\newline\urlprefix\url{https://doi.org/10.1038/ncomms7293}

\bibitem{thomas2022point}
D.~Thomas, Y.~Asiri, N.~Drummond,
  \href{https://doi.org/10.1103/PhysRevB.105.184114}{Point defect formation
  energies in graphene from diffusion quantum monte carlo and density
  functional theory}, Phys. Rev. B 105~(18) (2022) 184114.
\newblock \href {https://doi.org/https://doi.org/10.1103/PhysRevB.105.184114}
  {\path{doi:https://doi.org/10.1103/PhysRevB.105.184114}}.
\newline\urlprefix\url{https://doi.org/10.1103/PhysRevB.105.184114}

\bibitem{kirchhoff2022electronic}
A.~Kirchhoff, T.~Deilmann, P.~Kr{\"u}ger, M.~Rohlfing,
  \href{https://doi.org/10.1103/PhysRevB.106.045118}{Electronic and optical
  properties of a hexagonal boron nitride monolayer in its pristine form and
  with point defects from first principles}, Phys. Rev. B 106~(4) (2022)
  045118.
\newblock \href {https://doi.org/https://doi.org/10.1103/PhysRevB.106.045118}
  {\path{doi:https://doi.org/10.1103/PhysRevB.106.045118}}.
\newline\urlprefix\url{https://doi.org/10.1103/PhysRevB.106.045118}

\bibitem{tromer2025dynamic}
R.~M. Tromer, \href{https://doi.org/10.1021/acs.jctc.5c00856}{Dynamic collision
  fingerprints {(DCF)}: Introducing a new descriptor linking lattice
  interactions to {2D} structural data signatures}, J. Chem. Theory Comput.
  21~(16) (2025) 8106--8118.
\newblock \href {https://doi.org/https://doi.org/10.1021/acs.jctc.5c00856}
  {\path{doi:https://doi.org/10.1021/acs.jctc.5c00856}}.
\newline\urlprefix\url{https://doi.org/10.1021/acs.jctc.5c00856}

\bibitem{drude1902elektronentheorie}
P.~Drude, \href{https://doi.org/10.1002/andp.19003060312}{Zur elektronentheorie
  der metalle}, Annalen der Physik 312~(3) (1902) 687--692.
\newblock \href {https://doi.org/https://doi.org/10.1002/andp.19003060312}
  {\path{doi:https://doi.org/10.1002/andp.19003060312}}.
\newline\urlprefix\url{https://doi.org/10.1002/andp.19003060312}

\bibitem{schneider-1983}
T.~Schneider, Classical statistical mechanics of lattice dynamic model systems:
  transfer integral and molecular-dynamics studies, in: Statics and Dynamics of
  Nonlinear Systems: Proceedings of a Workshop at the Ettore Majorana Centre,
  Erice, Italy, 1--11 July, 1983, 1983, pp. 212--241.

\bibitem{hope2020linear}
T.~M. Hope, \href{https://doi.org/10.1016/B978-0-12-815739-8.00004-3}{Linear
  regression}, in: A.~Mechelli, S.~Vieira (Eds.), Machine learning, Elsevier,
  2020, pp. 67--81.
\newblock \href
  {https://doi.org/https://doi.org/10.1016/B978-0-12-815739-8.00004-3}
  {\path{doi:https://doi.org/10.1016/B978-0-12-815739-8.00004-3}}.
\newline\urlprefix\url{https://doi.org/10.1016/B978-0-12-815739-8.00004-3}

\bibitem{quinlan1986induction}
J.~R. Quinlan, \href{https://doi.org/10.1007/BF00116251}{Induction of decision
  trees}, Machine Learning 1~(1) (1986) 81--106.
\newblock \href {https://doi.org/https://doi.org/10.1007/BF00116251}
  {\path{doi:https://doi.org/10.1007/BF00116251}}.
\newline\urlprefix\url{https://doi.org/10.1007/BF00116251}

\bibitem{chen2016xgboost}
T.~Chen, C.~Guestrin, \href{https://doi.org/10.1145/2939672.293978}{{XGBoost}:
  A scalable tree boosting system}, in: Proceedings of the 22nd acm sigkdd
  international conference on knowledge discovery and data mining, Association
  for Computing Machinery, New York, NY, USA, 2016, pp. 785--794.
\newblock \href {https://doi.org/https://doi.org/10.1145/2939672.293978}
  {\path{doi:https://doi.org/10.1145/2939672.293978}}.
\newline\urlprefix\url{https://doi.org/10.1145/2939672.293978}

\bibitem{shi2021high}
X.~Shi, S.~Li, J.~Li, T.~Ouyang, C.~Zhang, C.~Tang, C.~He, J.~Zhong,
  \href{https://doi.org/10.1021/acs.jpclett.1c03193}{High-throughput screening
  of two-dimensional planar sp$^2$ carbon space associated with a labeled
  quotient graph}, J. Phys. Chem. Lett. 12~(47) (2021) 11511--11519.
\newblock \href {https://doi.org/https://doi.org/10.1021/acs.jpclett.1c03193}
  {\path{doi:https://doi.org/10.1021/acs.jpclett.1c03193}}.
\newline\urlprefix\url{https://doi.org/10.1021/acs.jpclett.1c03193}

\bibitem{ong2013python}
S.~P. Ong, W.~D. Richards, A.~Jain, G.~Hautier, M.~Kocher, S.~Cholia,
  D.~Gunter, V.~L. Chevrier, K.~A. Persson, G.~Ceder,
  \href{https://doi.org/10.1016/j.commatsci.2012.10.028}{Python materials
  genomics (pymatgen): A robust, open-source python library for materials
  analysis}, Comput. Mater. Sci. 68 (2013) 314--319.
\newblock \href
  {https://doi.org/https://doi.org/10.1016/j.commatsci.2012.10.028}
  {\path{doi:https://doi.org/10.1016/j.commatsci.2012.10.028}}.
\newline\urlprefix\url{https://doi.org/10.1016/j.commatsci.2012.10.028}

\end{thebibliography}
	
\end{document}